\documentstyle[12pt,aps]{revtex}

\begin{document}

\title{{\small International Conference on Theoretical Physics
Paris, UNESCO, 22-27 July 2002} \\
\vskip 0.5cm
Incomplete information and correlated electrons}

\author{Qiuping A. Wang}
\address{Institut Sup\'erieur des Mat\'eriaux du Mans,\\ 44, Avenue F.A.
Bartholdi, 72000 Le Mans, France}

\maketitle

\begin{abstract}

Although G\"odel's incompleteness theorem made mathematician recognize that no axiomatic system could
completely prove its correctness and that there is an eternal hole between our knowledge and the world, and in
spite of the work of Poincar\'e of about 100 years ago and the further development of the theory of chaos, the
dream of man to conquer nature and to know everything about nature refuse to die away. Physicists continue this
ambition in working so far on the approaches based on the hypothesis to completely or approximately know the
systems of interest. In this paper, however, I review the recent development of a different approach, a {\it
statistical theory based upon the notion of incomplete information}. Incomplete information means that, with
complex systems whose interactions cannot be completely written in its hamiltonian or whose equation of motion
does not have exact solution, the information needed to specify the systems is not completely accessible to us.
This consideration leads to generalized statistical mechanics characterized by an incompleteness parameter
$\omega$ which equals unity when information is complete. The mathematical and physical bases of the
information incompleteness are discussed.

The application of the concomitant incomplete fermion statistics to correlated electron systems is reviewed. By
comparison with some numerical results for correlated electron systems, it is concluded that, among several
other generalizations of Fermi-Dirac distribution, only the incomplete one is suitable for describing this kind
of systems. The extensive incomplete fermion distribution $n=1/\{exp[\omega(e - e_f)/k_BT]+1 \}$ gives very
good description of weakly correlated electrons with about $0.003<\omega<1$, the normalization index in
$\sum_ip_i^\omega=1$ where $p_i$ is probability distribution. On the other hand, the nonextensive fermion
distribution, $n=1/ \{ [1+(\omega-1)(e-e_f)/k_BT]^{\omega/(\omega - 1)}+1 \}$, does not show weak correlation
behaviors of electrons and is only suitable to describe strong correlated heavy fermion systems showing strong
increase of Fermi momentum with increasing correlations for $0<\omega<1$.
\end{abstract}


\section{Introduction}
As the study of complexity advanced, scientists have realized that chaotic and fractal behaviors were
ubiquitous in nature and the simple phenomena described by deterministic or quasi-deterministic\cite{Quasi}
physical sciences considering only simple interactions or predictable linear behaviors were only a few special
or accidental cases. It was also realized that patching up was fundamentally useless within the conventional
physics theories that break down once applied to complex systems having long range interactions or showing
nonlinear behavior related to chaotic or fractal phase space structure. Generalization of these theories would
be necessary. Driven by the increasing knowledge about chaos and fractals, the attempt of generalization has
been rapidly focused on the problems relative to information and statistics theory
\cite{Reny66,Hilb94,Ruelle91,Thuan,Complex,Tsal88,Wang01,Wang01a}. The development of the nonextensive
statistical mechanics (NSM)\cite{Tsal88,Wang01,Cura91,Tsal99,Penni}, among others\cite{Wang02d}, is a good
example of this tendency in physics.

Though considered by some to have a weak point due to the lack of clear physical significations of its
generalization parameter $q$, the probability distributions of NSM has been proved to be surprisingly useful
for describing complex systems having long term interactions or correlations for which Boltzmann-Gibbs
statistics (BGS) is no more valid. NSM generalizes BGS with a distribution function called $q$-exponential
given by $exp_q(x)=[1+(1-q)x]^{1/(1-q)}$. The latter is the inverse function of a generalized logarithm
$\ln_q(x)=\frac{x^{1-q}-1}{1-q}$ which can be used as a generalization of Hartley logarithmic information
measure to obtain the $q$-entropy $S_q=-k\frac{1-\sum_ip_i^q}{1-q}$ $(q \in R)$\cite{Wang01,Yama01} proposed by
Tsallis\cite{Tsal88}. When $q=1$, These above two generalized functions become the usual ones and the
$q$-entropy becomes Shannon one.\footnote{From now on, the parameter $q$ will be replaced by $\omega$ and
Tsallis entropy by $S_\omega=-k\frac{\sum_ip_i-\sum_ip_i^\omega}{1-\omega}$. The above generalized functions
will be called $\omega$-exponential and $\omega$-logarithm. I make this replacement for the simple reason that,
though it often gives similar forms of functions as $q$, $\omega$ defined in the framework of the theory I
review here does not have the same physical content as the parameter $q$ in Tsallis version of NSM. So I prefer
to use $\omega$ to avoid confusions. By definition, $\omega$ has clear physical meaning as the reader will find
in this paper.}

In the present paper, I will review our recent efforts to find consistent foundation for NSM distribution
functions and to give satisfactory answers to some fundamental questions. These efforts are based on a notion
which is both new and old : {\it incomplete information}\cite{Wang01,Wang01a}. New because scientists always
claimed, in constructing physics theories, that their theories contain all necessary information for specifying
the systems under consideration. This is the case of all the conventional physical theories : from Newtonian to
quantum physics, in passing by Einstein, Boltzmann and Shannon (certainly, a theory containing only partial
information about the system of interest is a little bit discouraging). Old because since the discovery of,
e.g., irrational numbers, mathematicians know that, within arithmetical system, they loss some information
about the world and that one could not know everything with infinite precision. In 1931, G\"odel
shown\cite{Ruelle91,Thuan,Complex} that mathematics system (or any axiomatic system) is incomplete in the sense
that within any such axiomatic system there is never sufficient information to prove all possible statements of
the theory\cite{Complex}. If a non negligible amount of information is not accessible to us, BGS theory has to
be modified. Incomplete information theory is a kind of modification (generalization) of BGS suggested by this
consideration as well as by some difficulties encountered within NSM in the last decade\cite{Wang01,Wang03}.

\section{Complete information assumption}
In this section, I will briefly review the well known information theory founded by Shannon et
al\cite{Shannon}. It should be remember that {\it information} about a real system is not our knowledge about
it. It is our ignorance. The ignorance of something to which we may have access. A mail address, as a state of
physical system, may be an information if we do not know it. More we know about a system, less there is
information in its description. So in a deterministic theory (e.g., classical mechanics), information is null.
In statistical theory, there is information because we ignore something so that we are not sure of the exact
state at any given moment of the system under consideration. So information can be related to the uncertainty
due to the ignorance or to the {\it probability} of finding the system at different states. It should be noted
that, as mentioned above, up to now, we always suppose that the information we address in any statistical
theory is complete or completely accessible. That is if we obtain it, we can answer all questions which can be
asked about the system. This certainty is reflected by the following postulate :
\begin{equation}                                            \label{1}
\sum_{i=1}^{v}p_i=1,
\end{equation}
where $v$ must be the number of all the possible states of the system under consideration. As a result, the
arithmetic average of $\xi$ is given by $\bar{x}=\sum_{i=1}^{v}p_ix_i$,

By some analysis of the information properties, it is supposed\cite{Reny66,Shannon} that the information is
given by the well know Hartley formula $\ln(N)$\cite{Hartley} needed to specify $N$ elements, or by
$\ln(1/p_i)$, the information needed to specify that an element will be found at the state $i$. If we perfectly
know all the $v$ possible states, then the complete information measure $I$ is given by averaging all
$\ln(1/p_i)$ :
\begin{equation}                                            \label{2}
I=\sum_{i=1}^{v}p_i\ln(1/p_i).
\end{equation}

It should be emphasize that the above definition of information or entropy needs the harsh condition that the
interactions in the system of interest are of short range or limited between the walls of the containers of
subsystems which are consequently independent of each other. To see this, it suffices to consider the
assumption of information additivity, i.e., for a system $C$ containing two subsystems $A$ and $B$, it is
supposed $I(C)=I(A)+I(B)$. This additivity is valid if and only if the information $I(C)$ needed in order to
specify simultaneously $A$ and $B$ is given by $\ln[N(A)N(B)]$ where $N(A)$ and $N(B)$ are respectively the
number of elements in $A$ and $B$. This is as if we had a system $C$ containing $N(A)N(B)$ elements. This
result needs that the states of the elements of $A$ do not depend on the states of $B$. In other words, these
is no interactions between the elements of $A$ and those of $B$. There may be interactions between the elements
on the walls of the containers of $A$ and $B$, but most of the elements inside $A$ and $B$ must be independent.
This is a case of short range interaction where we have not only additive information or entropy, but also
additive energy and other extensive thermodynamic variables.

I would like to recall in passing here that the total information $\ln[N(A)N(B)]$ implies
\begin{equation}                                            \label{3}
p_{ij}(C)=p_i(A)p_j(B)
\end{equation}
where $p_{ij}(C)$ is the probability that the composite system $C$ is at the product state $ij$ when $A$ is at
the state $i$ with probability $p_i(A)$ and $B$ at $j$ with $p_j(B)$. Eq.(\ref{3}) symbolizes the independence
of the noninteracting subsystems having additive physical quantities. But for interacting subsystems, it
symbolizes totally different physical reality. This product law has been widely employed and discussed in the
last decade in connection with equilibrium and many body
problems\cite{Guer96,Abe01a,Buyu93,Wang02e,Wang02b,Wang02a,Wang02c} and caused much confusion within NSM because
it paradoxically independence of subsystems and additive energy for nonextensive interacting systems. Very
recently, we shown that Eq.(\ref{3}) was nothing but {\it the consequence of the existence of thermodynamic
equilibrium in interacting systems described by $\omega$-entropy and did not need independence of the
subsystems}. This conclusion allows to exactly define equilibrium parameters such as temperature, pressure and
chemical potential for nonextensive systems and to obtain the exact one body quantum
distributions\cite{Wang02e,Wang02b,Wang02a,Wang02c}.

According to above discussions, we can say that, if there are long range interactions between $A$ and $B$, the
information about $C$ will be different from $\ln[N(A)N(B)]$ because the elements are correlated and can no
more occupy their states independently. According to the nature of the correlation, there may be {\it more or
less} information than in the noninteracting case. In general, we should write $I(C)=I(A)+I(B)+f[I(A),I(B)]$, a
case treated by NSM. Now Eq.(\ref{3}) becomes questionable, yet it is a crucial relationship for any
statistical mechanics, for it's applications to many-body systems and it's thermodynamics connection. The
reader will find detailed discussions on this issue below.

\section{Complexity and mathematics}
Certainly, complete information is possible whenever all possible states are well known so that we can count
them to carry out the calculation of probability and information. In physics, this requires that we can find the
{\it exact hamiltonian} and also the {\it exact solutions of the equation of motion} to know all the possible
states and to obtain the exact values of physical quantities dependent on the hamiltonian. The reader will see
that these two ``exact" conditions of complete information are almost impossible to satisfy.

Let us begin by asking some questions about the mathematical basis of physical theory.

What is the A basic field of mathematics is the classical arithmetic. From the epistemological point of view,
arithmetic is a theory based on a model of world resulted from the direct intuition of human beings. This is a
simple model for fragmented world containing only isolated, distinct and independent parts. So you have $1=1$,
$1+1=2$ and a series of rules, theorems and generalizations. No matter how complicated are the immense
mathematical constructions developed from arithmetic, their validity is always limited by these initial
conditions imposed by the crude data of our senses and direct intuition. Indeed, our senses, luckily, have the
capability of filtering the complex world into separated and discernible parts. If not, scientific knowledge
would be impossible. But these harsh constraints imposed by this filtration, as claimed by
Poincar\'e\cite{Poincare}, should not be forgotten. We have to ask the following question : how far he can go
with the concepts formed through the filtration in the real messy world or complex systems including
interacting, entangled and overlapped parts, especially when the interactions can no more be neglected.

So in some sense, it can be said that mathematics is an approximate theory containing finite amount of
information about the world which is surely incomplete because some information is lost by our senses through
the formation of the axioms. Any formation of axiomatic systems is necessarily made through a kind of
filtration of the world. The results of the filtration are not wrong, but they are only partially true.
Something about the connection of different parts of the world is rejected by the filtration. In my opinion,
this is $why$ axiomatic systems, as stated by the incompleteness theorem of G\"odel, inevitably fail to prove
some statements, especially those about their axioms. There is no enough information for that. The missing
information is just what rejected by the formation of axioms.

A mathematician is rather interested by the coherence of his logical systems based on axioms. He may put aside
the missing information and work within the logical systems without being connected to physical reality. But
for a physicist, the connection of his theory to the outside world is the most important thing he mind. He
possibly ask : My physical theory is in fact an application of a incomplete mathematical theory. If the
information I am handling is not complete, how can I apply it to the world whose description probably needs
more information?

In what follows, we will try to answer this question in recognizing that the incompleteness of all axiomatic
systems discovered by G\"odel has put an end to the ambition of establishing physical theories containing or
capable of treating complete information about any system in the world. In this sense, any physics theory is
incomplete by definition. This is the very reason for the introduction of ``incomplete information" into
statistical physics. This introduction needs in addition other considerations I am presenting below.

\section{Complexity and incomplete information}
Now let us look at the information problem from the physical viewpoint. I will try to show that, due to the
omnipresent $complexity$ in the world, we cannot have access to all the necessary information for complete
description of a system. Here ``complexity" means that the systems show nonlinear behaviors which are extremely
sensible to initial conditions and unpredictable. This is the famous $chaos$ observed almost everywhere in the
world\cite{Hilb94,Ruelle91,Thuan,Complex}.

A complex system is not necessarily a complicated system with a large number of freedoms. A one dimensional
oscillator with well known nonlinear interaction (with potential $\propto x^4$, for example) or a three body
system with gravitation ($\propto 1/r$) can behave chaotically. These two cases are just very good examples of
the impossibility of the two ``exact" conditions of complete information mentioned above. In the case of the
three body problem, we know (at least we believe that we know) the exact interaction of the system (Newtonian
gravitation). But Poincar\'e showed that the exact and predictable solution of the equation of motion was not
possible\cite{Ruelle91,Thuan}. There are in fact infinite number of periodic and aperiodic solutions. The
movement is chaotic and unpredictable and the attractors of the chaotic structures formed by the trajectories
in phase space are fractal. This means that we never know all possible states of the system and that complete
information treatment becomes impossible. We even have to redefine probability distribution in order to
calculate it in chaotic or fractal phase space.

Above conclusion is for hamiltonian systems whose interactions is {\it \`a priori} well known. When the
hamiltonian cannot be exactly written, the situation is more complicated. Even the exact and predictable
solutions of equation of motion are not complete due to the incomplete hamiltonian. This may happen if, for a
isolated closed system, the interactions are too complex to be written, or, for a system with simple
interactions, the effects of the external perturbations are not negligible. Sometimes negligible perturbations
may have drastic consequences if the system is sensitive to initial conditions. In this sense, the omission of
small interactions may make enormous information unaccessible to the theory. This incompleteness due to
neglected interactions simply adds to the incompleteness mentioned previously.

In any case, complete information description of complex systems is only a science fiction. Although we cannot
say that all these systems have chaotic or fractal nature, a common feature of them is that {\it a part of
their phase space is unknown so that complete and exhaustive exploit of the phase space is impossible}. The
calculable information is inevitably limited by this incompleteness of knowledge. That is evident. The
treatments of these systems based on the assumption of complete information and probability distribution are
not well founded.  They are legitimate only when unaccessible part of the information is negligible with
respect to the accessible information and to the desired precision of observation or theoretical description.

In what follows, we will try to introduce the notion of incompleteness of information into physics through
statistical method. It was with this method that man began to overcome the obstacle of his limited knowledge in
supposing, on the basis of Newtonian or quantum mechanics, that the missing knowledge (information) is
mathematically accessible or, equivalently, that the calculated probability must sum to one. Now if we say that
we cannot have access to every information we need or to every point of the phase space, a serious impact on
the normalization of probability, the very first stone in the construction of statistics, will be inevitable.

\section{Chaos and incomplete probability distributions}
\subsection{Incomplete normalization}
What can we do for probability and information calculation if we do not know how many states the system of
interest has? When we deal with a chaotic system having fractal attractor in phase space\cite{Hilb94}, it is as
if we toss a coin which often comes down, neither tails nor heads, but standing on the side without, in
addition, being observed. All calculations based on Eq.(\ref{1}) with $v=2$ would lead to aberrant results
because we have now $\sum_{i=1}^{v}p_i=Q\neq 1$. In this case, $p_i$ is referred to as {\it incomplete
distribution}\cite{Reny66} and $Q$ is {\it a constant depending on and characterizing the incompleteness} of
the system and provides {\it a possible key to introduce incompleteness of information into physics theory}. It
should be supposed $Q=1$ if information is complete.

The philosophy of incomplete information theory we developed is to keep the methods of classical complete
probability theory for incomplete information or probability distribution by introducing empirical parameters
in order to characterize the incompleteness. This is just the same methodology as in the theory of chaos or
fractals introducing fractal dimension to characterize the structures of space time. In this sense, we can
refer to the parameter $\omega$ introduced below as {\it incompleteness parameter}.

First of all, we need a ``normalization" for incomplete distribution $p_i$ in order to take advantage of the
conventional probability theory. This is an occasion to introduce a parametrization function $F_\omega$ and to
write
\begin{equation}                                            \label{4}
\sum_iF_\omega(p_i)=1
\end{equation}
which can be called generalized or {\it incomplete normalization}. $F_\omega$ should depend on the nature of
the system and become identity function whenever information is supposed complete ($Q=1$). The arithmetic
average should now be given by $\bar{x}=\sum_iF_\omega(p_i)x_i$. $F_\omega$ can be determined if the
information measure and the distribution law are given. For example, with Hartley information measure and
exponential distribution, $F_\omega$ can be showed to be identity function\cite{Wang01a}. In general, by
entropy maximization through the functional
\begin{equation}                                            \label{5}
\delta [\sum_iF_\omega(p_i)I(p_i)+\beta\sum_iF_\omega(p_i)x_i]=0
\end{equation}
we get :
\begin{equation}                                            \label{6}
\frac{\partial \ln F_\omega(p_i)}{\partial p_i}=\frac{\partial I/\partial p_i}{I+\beta f_\omega^{-1}(p_i)}
\end{equation}
or $F_\omega(p_i)=C\exp[\int\frac{\partial I/\partial p_i}{I+\beta f_\omega^{-1}(p_i)}dp_i]$ where $\beta$ is
the multiplier of Lagrange connected to expectation, $I(p_i)$ is the information measure, $p_i=f_\omega(x_i)$
the distribution function depending on the parameter $\omega$, $C$ the normalization constant of $F_\omega$.

\subsection{Incomplete normalization of NSM}

In my previous papers\cite{Wang01,Wang01a}, in order to find coherent foundation for $\omega$-exponential
distribution on the basis of $\omega$-logarithm information measure, $F_\omega(p_i)=p_i^\omega$ was postulated.
So that
\begin{equation}                                            \label{7}
\sum_ip_i^\omega=1.
\end{equation}
In what follows, I will try to show that the conjecture of power law incomplete normalization in the previous
section is inevitable in a chaotic or fractal space time.

For the sake of simplicity, let us consider a phase space in which the trajectory of a chaotic system forms a
simple self-similar fractal structure, say, Sierpinski carpet (Figure 1). This means that the state point of
the system can be found only on the black rectangular segments whose number is $W_k=8^k$ at $k^{th}$ iteration.
Hence the total surface at this stage is given by $S_k=W_k s_k$ where $s_k=l_0/3^k$ is the surface of the
segments at $k^{th}$ iteration and $l_0$ the length of side of the square space at $0^{th}$ iteration. If the
segments do not have same surface, we should write $S_k=\sum_{i=1}^{W_k}s_k(i)$. We suppose that the density of
state is identical everywhere on the segments and that the distribution is microcanonical, so that the
probability for the system to be in the $i^{th}$ segment may be defined as usual by $p_i=s_k(i)/S_k$. This
probability is obviously normalized. The problem is that, as discussed in \cite{Hilb94}, $S_k$ is an indefinite
quantity as $k\rightarrow \infty$ and, strictly speaking, can not be used to define exact probability
definition. In addition, $S_k$ is not differentiable and contains inaccessible points. Thus the probability
defined above makes no sense.

Alternatively, the probability may be reasonably defined on a integrable and differentiable support, say, the
Euclidean space containing the fractal structure. To see how to do this, we write
$S_k=l_0^2(\frac{1}{3^k})^{d-d_f}$ for identical segments or, for segments of variable size,
\begin{equation}                                \label{8}
\sum_{i=1}^{W_k}[\frac{s_k(i)}{S_0}]^{d_f/d}=1
\end{equation}
where $S_0=l_0^d$ (here $d=2$ for Sierpinski carpet) a characteristic volume of the fractal structure embedded
in a $d$-dimension Euclidean space, $d_f=\frac{\ln n}{\ln m}$ is the fractal dimension, $n=8$ the number of
segments replacing a segment of the precedent iteration and $m=3$ the scale factor of the iterations. The
microcanonical probability distribution at the $k^{th}$ iteration can be defined as $p_i=\frac{s_k(i)}{S_0}$ so
that $\sum_{i=1}^{W_k}p_i^{d_f/d}=1$ which is just Eq.(\ref{7}) with $\omega=d_f/d$. The conventional
normalization $\sum_{i=1}^{W_k}p_i=1$ can be recovered when $d_f=d$.

It should be noticed that, in Eq.(\ref{8}), the sum over all the $W_k$ segments at the $k^{th}$ iteration does
not mean the sum over all possible states of the system under consideration. This is because that the segment
surface $s_k(i)$ does not represent the real number of state points on the segment which, as expected for any
self-similar structure, evolves with $k$ just as $S_k$. So at any given order $k$, the complete summation over
all possible segments is not a complete summation over all possible states. But in any case, whatever is $k$,
Eq.(\ref{8}) and $\sum_{i=1}^{W_k}p_i^\omega=1$ always holds for $\omega=d_f/d$.

In this simple case with self-similar fractal structure, the incompleteness of the normalization Eq.(\ref{7})
is measured by the parameter $\omega=d_f/d$. If $d_f>d$, there are more state points than $W_k$, the number of
accessible states at given $k$. If $d_f<d$, the number of accessible states is less than $W_k$. When $d_f=d$,
the summation is complete at any order $k$, corresponding to complete information calculation.

\section{Incompleteness parameter $\omega$}
Here I will discuss in a detailed way the incompleteness parameter $\omega$ and its physical meanings.
Incomplete statistics gives to the empirical parameter $\omega$ a clear physical signification : {\it measure of
the incompleteness of information or of chaos}. Let us illustrate this by the simple case of self-similar
fractal phase space with segments of equal size.

\subsection{$\omega$ and phase space expansion}
As discussed in the case of chaotic phase space, $\omega=\ln n/d\ln m$ gives a measure of the incompleteness of
the state counting in the $d$-dimension phase space. $\omega=1$ means $d_f=d$ or $n=m^d$. In other word, at the
$k^{th}$ iteration, a segment of volume $s_k$ is completely covered (replaced) by $n$ segments of volume
$s_{k+1}=s_k/m^d$. So the summation over all segments is equivalent to the sum over all possible states, making
it possible to calculate complete information.

When $\omega>1$ (or $\omega<1$), $n>m^d$ (or $n<m^d$) and $s_k$ is replaced by $n$ segments whose total volume
is more (or less) than $s_k$. So there is expansion (or negative expansion) of state volume when we refine the
phase space scale. An estimation of this expansion at each scale refinement can be given by the ratio
$r=\frac{ns_{k+1}-s_k}{s_k} =\frac{n}{m^d}-1 =(\frac{1}{m^d})^{1-\omega}-1 =
(\omega-1)\frac{(m^d)^{\omega-1}-1}{\omega-1}$. $r$ describes {\it how much unaccessible states increase} at
each step of the iteration or of the refinement of phase space. The physical content of $\omega$ is clear if we
note that $\omega>1$ and $\omega<1$ correspond to an expansion ($r>0$) and a negative expansion ($r<0$),
respectively, of the the state volume at each step of the iteration. When $\omega=0$, we have $d_f=0$ and $n=1$,
leading to $r=\frac{1}{m^d}-1$. The iterate condition $n\geq 1$ means $\omega\geq 0$, as proposed in references
\cite{Wang01}. $\omega<0$ is impossible since it means $d_f<0$ or $n<1$ which obviously makes no sense. We can
also write : $\omega-1=\ln(r+1)/\ln(m^d) =\ln(ns_{k+1}/s_k)/\ln(m^d)$, which implies that it is the difference
$\omega-1$ which is a direct measure of the state space expansion through the scale refinement.

\subsection{$\omega$ and information growth}

The expansion of the state volume of a system in its phase space during the scale refinement should be
interpreted as follows : the extra state points $\Delta=ns_{k+1}-s_k$ acquired at $(k+1)^{th}$ order iterate
with respect to $k^{th}$ order are just the number of unaccessible states at $k^{th}$ order. $\Delta>0$ (or
$\Delta<0$) means that we have counted less (or more) states at $k^{th}$ order than we should have done.
$\Delta$ contains the {\it accessible information gain} (AIG) through the $(k+1)^{th}$ iterate.

To illustrate the relation between this ``hidden information" and the parameter $\omega$, let us first consider
the Hartley logarithm information in the simple case where {\it the distribution is microcanonical and
scale-invariant}\cite{Invariant}. At the iterate of order $k$, the average information contained on $s_k$ is
given by $I_k=\int_{s_k} p^\omega\ln(1/p)ds$. At $k+1$ order, $I_{k+1}=\int_{ns_{k+1}} p^\omega\ln(1/p)ds$.
Hence AIG is just $\Delta I=I_{k+1}-I_k=\int_{(ns_{k+1}-s_k)} p^\omega\ln(1/p)ds =\sigma_I\Delta$, where
$\sigma_I=p^\omega\ln(1/p)$ is the information density or the average information carried by each state. The
relative AIG is given by $\Delta I/I_k=r=(1-\omega)\frac{(1/m^d)^{1-\omega}-1}{1-\omega}$ which is independent
of scale but dependent on scale changes. For given scaling factor $m$, the magnitude of $\Delta I$ or $r$
increases with increasing difference $|1-\omega|$. The sign of $r$ (or AIG) was discussed earlier. For given
$\omega$, $|\Delta I|$ increases with decreasing scaling. For $\omega=1$ or $m=1$, there is no information gain,
corresponding to the case of complete information.

According to the relationship $\omega=d_f/d$ and the above discussions, it can be concluded that the
incompleteness parameter $\omega$ may be considered as a measure of chaos. Certainly this is a conclusion on the
basis of simple models and the relation between $\omega$ and the degree of chaos or fractal may be more
complicated with more complex chaos and fractals, but it is consequent to say that more a system is chaotic,
more its information is incomplete and more $\omega$ is different from unity.

\section{Nonadditive incomplete distributions}
To get the nonextensive distribution in $\omega$-exponential as mentioned above, we can maximize the entropy
$S_\omega=-k\frac{\sum_ip_i-\sum_ip_i^\omega}{1-\omega}$ \cite{Wang01,Wang01a} according to the Jaynes
principle\cite{Jaynes} with the constraints $U=\sum_ip_i^\omega E_i$ and $N=\sum_ip_i^\omega N_i$ for
grand-canonical ensemble, where $U$ is the internal energy, $N$ the average particle number, $E_i$ the energy
and $N_i$ the particle number at the state $i$ of the system. We obtain :
\begin{equation}                                    \label{9}
p_i=\frac{[1-(1-\omega)\beta(E_i-\mu N_i)]_\dag^\frac{1}{1-\omega}}{Z}.
\end{equation}
where $Z^\omega=\sum_{i}^v[1-(1-\omega)\beta(e_i-\mu N_i)]_\dag^\frac{\omega}{1-\omega}$. $[x]_\dag=x$ if $x>0$
and $[x]_\dag=0$ otherwise. $\beta$ is the inverse temperature and $\mu$ the chemical potential. This
distribution function has been proved particularly useful for systems showing non gaussian distribution
functions (for detailed information, see \cite{Tsal99} and references there-in). Considering Eq.(\ref{3}), the
product probability law at thermodynamic equilibrium, the one-particle distribution from Eq.(\ref{9}) can be
rewritten as $p_k=\frac{[1-(1-\omega)\beta(e_k-\mu)]_\dag^\frac{1}{1-\omega}}{z}$ where $e_k$ is the energy of
one particle at the state $k$ and $z_n^\omega=\sum_k[1-(1-\omega)\beta(e_k-\mu)]_\dag^\frac{\omega}{1-\omega}$
is the one-particle partition function.

As shown in \cite{Wang02a}, the above one-particle distribution can be recast into exponential form as follows
\begin{equation}                                    \label{10}
p_k = \frac{1}{z}[1-(1-\omega)\beta'e_k]^\frac{1}{1-\omega}[1+(1-\omega)\beta'\mu']^\frac{1}{1-\omega}
=\frac{1}{Z}e^{-\beta'(\epsilon_k-\nu)}
\end{equation}
where $\beta'=\frac{\beta}{1-(q-1)\beta\mu}$, $\mu'=\mu[1-(q-1)\beta\mu]$ which imply $\beta'\mu'=\beta\mu$,
$\nu=\frac{\ln[1+(1-q)\beta'\mu']}{(1-q)\beta'}$ and $\epsilon_k=\frac{\ln[1+(q-1)\beta' e_k]}{(q-1)\beta'}$.
The exponential distribution Eq.(\ref{10}) makes it possible to straightforwardly obtain the {\it exact quantum
distribution}\cite{Wang02a} (EQD) given by
\begin{eqnarray}                                        \label{11}
\bar{n}_k=\frac{1}{e^{\omega\beta'(\epsilon_k-\nu)}\pm1}
=\frac{1}{[1+(\omega-1)\beta(e_k-\mu)]^{\frac{\omega}{\omega-1}}\pm1},
\end{eqnarray}
where $\bar{n}_k$ is the occupation number of the one-particle state $k$. "+" is for fermions and "-" for
bosons. These distribution can be compared to the approximate quantum distributions (AQD) of NSM\cite{Buyu93}
$\bar{n}_k =\frac{1}{e^{\beta'(\epsilon_k-\nu)}\pm1}=\frac{1}{[1+(q-1)\beta(e_k-\mu)]^{\frac{1}{q-1}}\pm1}$
given within a factorization approximation using additive energy. At first glance, EQD and AQD are not very
different from each other if we put $\omega=q$. But Figure 2 shows that they are two very different
distributions. AQD remains approximately the same as the conventional Fermi-Dirac distribution for whatever $q$
value. So its Fermi energy $e_f$ is almost constant with changing $q$. On the contrary, EQD changes drastically
with $\omega$. The Fermi energy $e_f$ shows a strong increase with decreasing $\omega$ up to two times
$e_{f_0}$ of the conventional Fermi-Dirac distribution when $\omega\rightarrow 0$. This $e_f$ increase has been
indeed noticed through numerical calculations for strongly correlated heavy electrons on the basis of
tight-binding Kondo lattice model\cite{Corelec2,Corelec3} as shown in Figure 3. Increasing correlation
corresponds to decreasing $\omega$ from unity (zero correlation). This implies that EQD based on incomplete
information has its merit in the description of heavy electron systems. Further investigation is needed to know
the connection between the correlation and the nonextensive parameter $1-\omega$.

\section{Additive incomplete distributions}
Although the nonextensive EQD accounts for an important aspect of correlated electrons, i.e., the correlation
induced Fermi energy increase, another important aspect of the weak correlation is missing in the description
of nonextensive EQD. This is the flattening of $n$ drop at $e_f$\cite{Corelec2,Corelec3,Corelec1,Corelec2a}.
That is the correlation, even at low temperature, drives electrons above $e_f$ so that the $n$ discontinuity
becomes less and less sharp as the correlation increases. Curiously, this flattening of $n$ discontinuity at
$e_f$ is completely absent in EQD of NSM. From Figure 2, we see that the sharp $n$ drop at $e_f$ is independent
of $\omega$ or correlations.

In what follows, I will present an additive incomplete statistical mechanics. It is assumed that the additive
Hartley information measure still holds. So with respect to the conventional Shannon information theory and
BGS, only the normalization is changed according to Eq.(\ref{7})\cite{Wang01a,Wang02d}. The additive incomplete
entropy is given by $S=k\sum_{i=1}^w p_i^\omega\ln(1/p_i)$. When $\omega\rightarrow 1$, $S$ is Shannon entropy,
which identifies $k$ to Boltzmann constant.

For {\it grand canonical ensemble}, the usual entropy maximization procedure leads to $p_i=e^{-\omega\beta
(E_i-\mu N_i)}/Z$ where partition function is given by $Z=\{ \sum_{i=1}^w e^{-\omega\beta (E_i-\mu N_i)}
\}^{1/\omega}$. For quantum particle systems, we have
\begin{equation}                                \label{12}
\bar{n}_k=\frac{1}{e^{\omega\beta (e_k-\mu)}\pm 1}.
\end{equation}

The fermion distribution given by Eq.(\ref{12}) is plotted in Figure 3 for different $\omega$ values in
comparison with some numerical simulation results. We note that IFD reproduces well the numerical results for
about $J<1$. When coupling is stronger, a long tail in the KLM distributions begins to develop at high energy.
At the same time, a new Fermi surface at $k=k_{f_0}+\pi/2=0.75\pi$ starts to appear and a sharp $n$ drop takes
place at the new Fermi momentum. At $J=4$, KLM distribution (x-marks) is very different from IFD (e.g.
$\omega=0.0011$). The solid line fitting better the $J=4$ KLM distribution is given by the incomplete
statistics version of fractional exclusion distribution
$(1/n-\alpha)^\alpha(1/n-\alpha+1)^{1-\alpha}=e^{\omega\beta(e-e_f)}$\cite{Wu01,Hald91} with $1/\alpha=0.85$
due to the KLM occupation number smaller than 0.5 at low momentum $k$.

\section{Conclusion}
Summing up, I have discussed the philosophical basis of incomplete information from both the viewpoints of
mathematical and physics. The information we deal with in scientific theories can not be complete in the sense
that a part of the information necessary for complete description of the system under consideration is not
accessible to our theory or knowledge. This part of information is rejected from scientific knowledge by the
formation of concepts, axioms and models. The amount of rejected information is particularly important for
complex systems having chaotic behaviors and fractal phase space. A parameterized normalization
$\sum_ip_i^\omega=1$ is proposed for this kind of systems, where $\omega$ is the incompleteness parameter
characterizing the inaccessibility of phase space points or of the information of the system. It also offers a
measure of the degree of chaos.

The wide drop in the fermion occupation number and the sharp cutoff of occupation number at $e_f$ showing
strong increase with increasing interaction can be interpreted by the nonextensive incomplete fermion
distribution with decreasing $\omega$ value. On the other hand, it fails to describe weak correlation effect on
electrons which is well accounted for by additive incomplete fermion distribution. But the additive
distribution does not show the sharp cutoff at $e_f$ when correlation is strong. This result suggests to
combine these two partially valid models to describe correlated electrons in a global way. Further results of
this current work will be presented in other papers of ours.

\acknowledgments : I acknowledge with great pleasure the useful discussions with J.P. Badiali, A. Le
M\'ehaut\'e, L. Nivanen, G. Kaniadakis and P. Quarati.

\newpage

{\Large Figure caption :}

\begin{figure}[p]
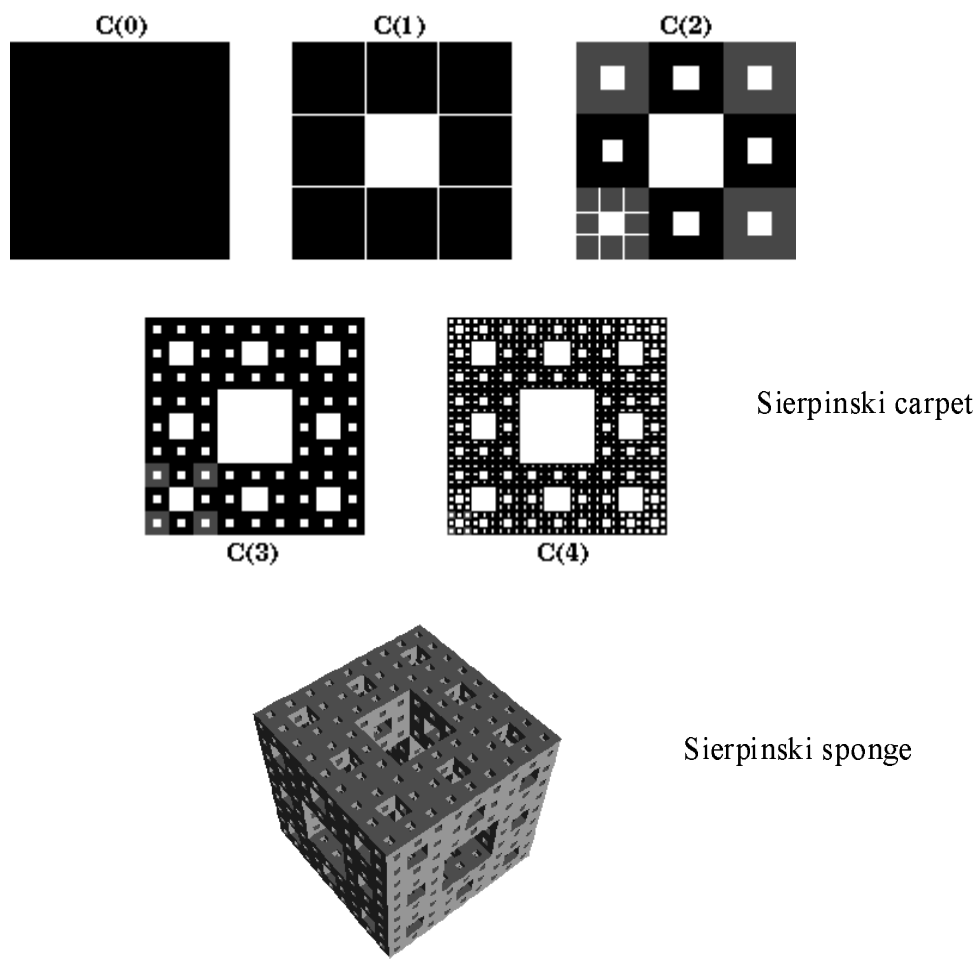
 \label{f1}
\caption{A simple model of fractal phase space in Sierpinski carpet (or sponge). At $k^th$ iteration, the side
of the squares (black or white) is $l_k=l_0/2^k$ and their number is $W_k=8^k$, $l_k$ being the length of the
side at at $0^{th}$ iteration. The total surface at $k^th$ iteration is $S_k=W_k s_k$ or  $W_k s_k/S_k=1$. The
classical probability definition by relative frequency of visits of each point by the system must be modified
because the total number of visits (propotional to black surface $S_k$ of the carpet) is no more a finite
quantity. (Construction of Sierpinski carpet. First iteration c(1) : removing the central square formed by the
straight lines cutting each side into three segments of equal size. Repeat this operation on the 8 remaining
squares of equal size and so on.)}
\end{figure}

\begin{figure}[p] \label{f2}
\caption{Nonextensive fermion distributions of AQD and EQD of incomplete statistical mechanics. AQD
distribution is only slightly different from that at $q=1$ (conventional Fermi-Dirac distribution) even with
$q$ very different from unity. But EQD changes drastically with decreasing $\omega$. As $\omega\rightarrow 0$,
the occupation number tends to 1/2 for all states below $e_f$ which increases up to 2 times $e_{f_0}$, the
conventional fermi energy at $T=0$.}
\end{figure}

\begin{figure}[p] \label{f3}
\caption{Comparison of additive incomplete fermion distribution (IFD, lines) with the numerical results
(symbols) of Eder el al on the basis of  Kondo lattice $t-J$ model (KLM) for different coupling constant $J$
[{\em Phys. Rev. B,} {\bf 55}(1997)6109]. In my calculations, the density of electrons is chosen to give
$k_{f_0}=0.25\pi$ in the first Brillouin zone. We note that IFD reproduces well the numerical results for about
$J<1$. When coupling is stronger, a long tail in the KLM distributions begins to develop at high energy.  At
the same time, a new Fermi surface at $k=k_{f_0}+\pi/2=0.75\pi$ starts to appear and a sharp $n$ drop takes
place at the new Fermi momentum. At $J=4$, KLM distribution (x-marks) is very different from IFD (e.g.
$\omega=0.0011$). The solid line fitting better the $J=4$ KLM distribution is given by the incomplete
statistics version of fractional exclusion distribution
$(1/n-\alpha)^\alpha(1/n-\alpha+1)^{1-\alpha}=e^{\omega\beta(e-e_f)}$ [Yong-Shi Wu, {\em Phys. Rev. Lett.,}
{\bf 73}(1994)922] with $1/\alpha=0.85$ due to the KLM occupation number smaller than 0.5 at low momentum $k$.}
\end{figure}


\begin{thebibliography}{99}
\bibitem {Quasi}
By quasi-deterministic theory, I mean conventional probabilistic sciences which, although incapable of giving
deterministic description as does Newtonian mechanics, give nevertheless a complete description based on a
perfect knowledge of all possible states of the system of interest. We are not sure of the state of the system
at any given moment, but we know that the system must be at one of the states in our register given by the
equation of motion.
\end{thebibliography}
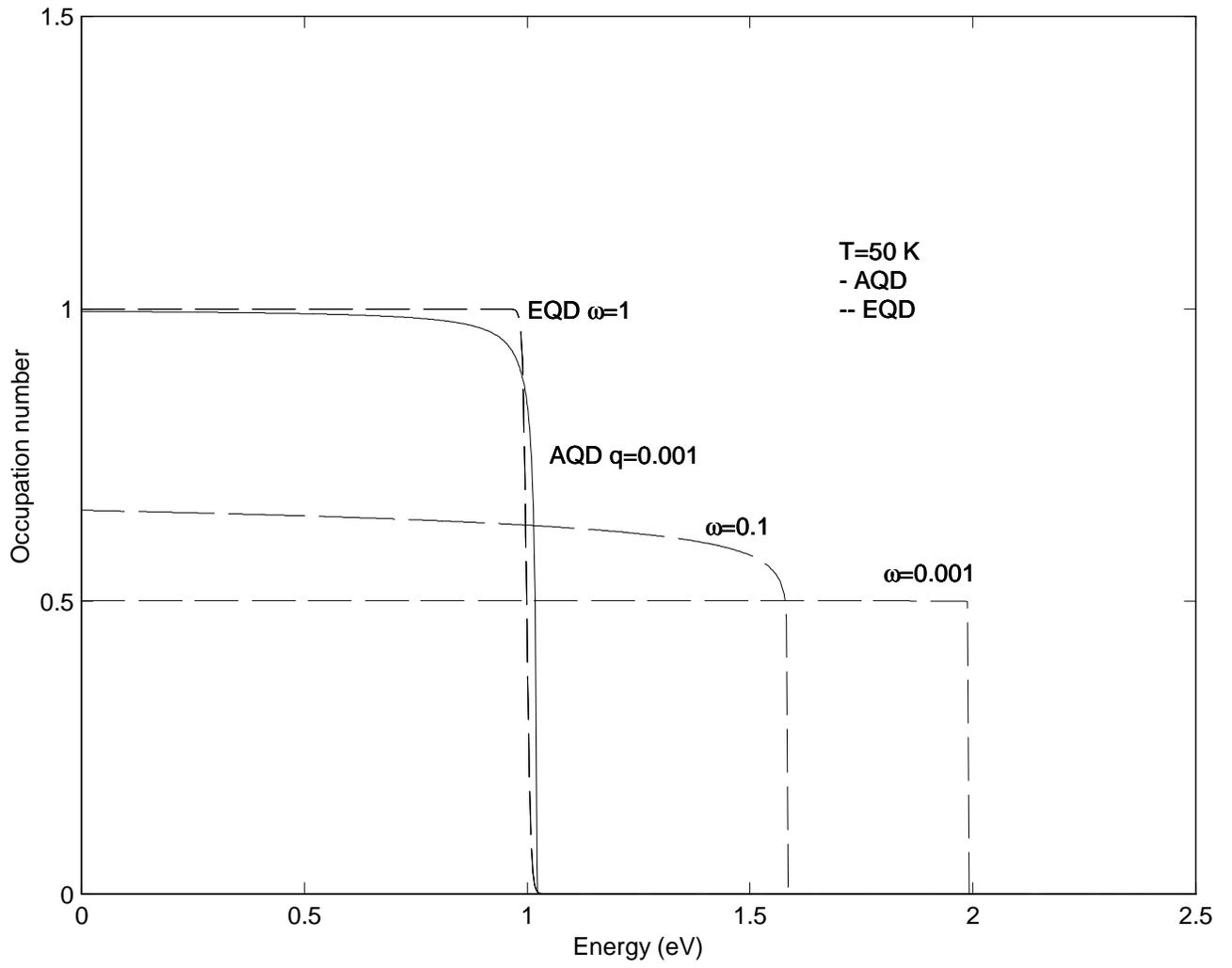
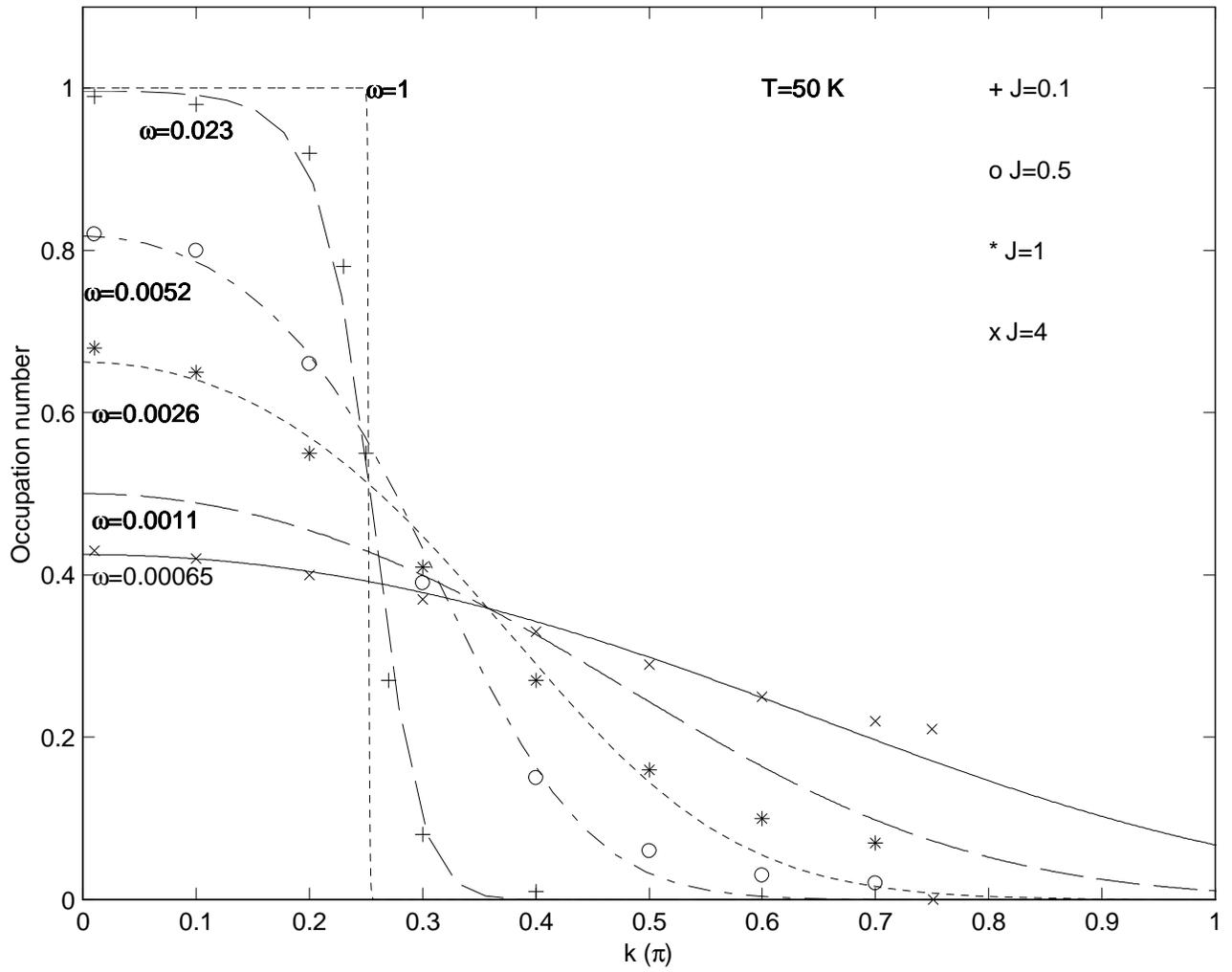
\end{document}